# The Rectified Second Law of Thermodynamics

Dor Ben-Amotz and J. M. Honig

Purdue University, Department of Chemistry, West Lafayette, IN 47907

**Abstract**

Equilibrium thermodynamics is combined with Jarzynski's irreversible work theorem to quantify the excess entropy produced by irreversible processes. The resulting rectified form of the second law parallels the first law, in the sense that it facilitates the experimental measurement of excess entropy changes resulting from irreversible work and heat exchanges, just as the first law quantifies energy changes produced by either reversible or irreversible work and heat exchanges. The general form of the rectified second law is further applied to a sub-class of quasi-static irreverisble (QSI) processes, for which all the thermodynamic functions of both the system and surroundings remain continuously well-defined, thus facilitating excess entropy measurements by integrating exact differential functions along QSI paths. The results are illustrated by calculating the mechanical and thermal excess entropy produced by the irreversible unfolding of an RNA molecule.

# 1. Introduction

The second law of thermodynamics is arguably the most enigmatic and provocative fundamental statement of relevance to engineering, physics, chemistry and biology. It is instructive to recall that the discovery of this profound principle of nature originally emerged from the practical concerns of nineteenth century industrialists who desired to increase the efficiency of steam engines.[1] Although the second law of thermodynamics may be stated in a myriad of ways, in the introduction to his famous paper *On the Equilibrium of Heterogeneous Substances*[2] Gibbs quoted Clausius' remarkably concise statement of the first and second laws of thermodynamics (as translated from German):

> The energy of the universe is constant.
> 
> The entropy of the universe seeks a maximum.

Thus, the second law may be summarized by the deceptively simple inequality, $\Delta S_{Univ} \geq 0$, pertaining to the increase in the entropy of the universe (or of any isolated entity) produced as a result of spontaneous (irreversible) processes. Although there is no question regarding the wide ranging implications of this and other statements of the second law, it is important to note that none of these statements can in themselves be used to quantify the excess entropy produced as the result of a given irreversible process.

Here we combining the results of nineteenth century thermodynamics with Chris Jarzynski's late twentieth century irreversible work theorem[3,4] (and its subsequent generalizations and implementations)[5-17] to rectify second law inequalities by converting them to an equalities which can be used to experimentally quantify irreversible entropy production.[18] The most general of the resulting expressions pertains to any irreversible process which may be represented by the combined performance of irreversible adiabatic work along generalized constraint displacement coordinates and irreversible heat exchange carried out with fixed constraints. Practical considerations limit the application of such expressions (either experimental or numerically) to molecular or mesoscopic processes, as can best be understood using very interesting and general reciprocal relations recently identified by Jarzynski.[5]

We further describe a broad class of quasistatic irreversible (QSI) processes for which the entropy produced as the result of generalized mechanical and/or thermal irreversibilities may be experimentally quantified. For simplicity, we assume that the surroundings behave like an ideal bath (which may or may not remain fully equilibrated with the system), and impose general conservation conditions (although these restrictions could readily be relaxed). Moreover, we illustrate the way in which Legendre transformations may be used to express irreversibly produced entropy as a function of any complete set of either extensive or intensive variables of the system (and surroundings).

The remainder of this paper is organized as follows. Section 2 summarized the first law of thermodynamics, with particular stress on the proper identification of work and heat exchanges as ensemble averages, which is critical to our subsequent analysis of the second law. Section 3 combines standard thermodynamic expressions with the Jarzynski equality[3] to produce a rectified form of the second law which may be used to quantify irreversible entropy production; these results follow and expand-on our recent *Physical Review Letter* pertaining to this subject.[18] Section 4 describes a broad class of quasistatic irreversible processes for which entropy production may be quantified in terms of any desired set of independent system variables (and the values of the intensive functions of an ideal bath). Section 5 contains a summary and discussion of the ways in which the above results may be applied, as exemplified by calculating the excess entropy produced as a result of the irreversible unfolding of an RNA molecule.

## 2. The First Law of Thermodynamics

The first law of thermodynamics asserts that the energy, $U$, of any system can only be changed by exchanging work, $W$, and/or heat, $Q$, with its surrounding.

$$\Delta U = W + Q \qquad (1)$$

Although this expression is in some respects self-evident, it's meaning and consequences are not free of subtleties. For example, all of the terms in eq 1 are understood to represent thermodynamic (average) quantities, and yet the amount of work and heat exchanged in particular realization of a process will in general depend on the specific micro-states of the system which are sampled during that process. Thus, the thermodynamic work, $W$, should be understood to represent the average value of the work, $w_i$, exchanged in many specific realizations of a process, weighted by the appropriate statistical mechanical probability density, $p_i$.

$$W \equiv \langle w \rangle = \lim_{n \to \infty} \left[ \frac{1}{n} \sum_{i=1}^{n} p_i w_i \right] \qquad (2a)$$

More specifically, a given process is defined by the associated constraint displacement(s), $X(t)$, and so the required average is obtained by repeating identical displacement(s) many times, in each case starting with a different initial microscopic configuration of the system (randomly picked from the corresponding initial equilibrium distribution). A thermodynamic heat exchange, $Q$, is understood to represent a similar average of particular heat exchanges, $q_i$, weighted by the same probability density.

$$Q \equiv \langle q \rangle = \lim_{n \to \infty} \left[ \frac{1}{n} \sum_{i=1}^{n} p_i q_i \right] \tag{2b}$$

Since $W$ and $Q$ are not state functions one may identify different paths ($A$, $B$, $C$…), each characterized by different constraint displacements, and each connecting the same initial and final states of a system. Thus, although each path may involve different $Q$ and $W$ exchanges, the first law requires all of their sums to be equal.

$$\Delta U = W_A + Q_A = W_B + Q_B = W_C + Q_C = \ldots \tag{3a}$$

It is also important to note that, although each of the above paths are assumed to start and end at the same two equilibrium states of the *system*, the initial and final states of the *surroundings* are in general path dependent.

Some of the above paths may be reversible while others are irreversible. The defining quality of any reversible process is that the system and surroundings remain arbitrarily close to internal equilibrium *and* the values of each of their intensive variables remain infinitesimally close to each other, to the extent allowed by the constraints imposed by the boundary between the system and it surrounding. For example, if the boundary is thermally conductive then the temperatures of the system and surroundings must remain nearly equal; if the boundaries are mechanically movable then the pressure of the system must remain infinitesimally close to that of the surroundings; if the boundaries are permeable to a given chemical species then the chemical potential of that species in the system and surroundings must remain infinitesimally close to each other.

Work exchanges may in general be expressed as an integral, $\int f dX$, of a generalized force, $f$, times a generalized displacement (constraint) variable, $X$. For example $f$ may be the force exerted on a piston while $X$ is the corresponding piston displacement. For a reversible process the generalized force tracks an intensive variable of the system (such as its pressure, $P$, or the chemical potentials of each compound, $\mu_i$, as further discussed in Section 4). For an irreversible process the system need not remain close to any equilibrium path, and so $f$ need not track a thermodynamic variable of either the system or surroundings.

The heat exchanged in a reversible process may be converted to an exact differential (state function) by introducing the inverse temperature as an integrating factor, such that $\oint \frac{\delta Q}{T} = 0$ (where the symbol $\delta$ is here reserved for infinitesimal *reversible* heat or work exchanges, and the integral is performed over any cyclic path). Since any heat exchanges observed from the perspective of the system are necessarily opposite in sign to those observed from the point of view of the surroundings, $\frac{\delta Q}{T} = -\frac{\delta Q_0}{T_0}$ (where the subscript 0 pertains to the surroundings, while the un-subscripted quantities pertain to the system). Moreover, if the surroundings behave like an ideal heat bath, as is typically assumed, then from the perspective of the bath, *any* heat exchange becomes indistinguishable from a reversible heat exchange. Note that an ideal bath is here defined as a bath which is sufficiently large (and well stirred) that all of its *intensive* thermodynamic functions may be treated as effectively constant (i.e. the temperature, pressure and chemical potentials of the bath are practically insensitive to any heat or work exchanges with the system).

Notice that eq. 3a requires that for any infinitesimal process,

$$dU = \delta W + \delta Q = \bar{\delta}W + \bar{\delta}Q \qquad (3b)$$

and so,

$$\bar{\delta}Q = \delta Q + \bar{\delta}W - \delta W = \delta Q + \bar{\delta}W_{diss} \qquad (4)$$

where the symbol $\bar{\delta}$ is used to indicate any infinitesimal work and heat exchanges (which may in general be irreversible), and $\bar{\delta}W_{diss} \equiv \bar{\delta}W - \delta W \geq 0$ is the dissipative work (and the last inequality follows from the second law, as further discussed below).

## 3. The Second Law of Thermodynamics

The exact differential, $\dfrac{\delta Q}{T}$, is equivalent to the change in the entropy of the system, $dS$, while the associated entropy change of the surroundings is $dS_0 = \dfrac{\delta Q_0}{T} = -\dfrac{\delta Q}{T}$. If the process producing a given heat exchange is not reversible, but the surroundings are assumed to behave like an ideal heat bath, then $dS_0 = -\dfrac{\bar{\delta}Q}{T}$. For a non-infinitesimal process the total entropy change of the system is $\Delta S = \int \dfrac{\delta Q}{T}$ (which implies that a reversible path is followed) while that of the ideal heat bath is $\Delta S_0 = -\int \dfrac{\bar{\delta}Q}{T_0} = -\dfrac{Q}{T_0}$.

The second law inequality may thus be represented as

$$\Delta S_{Univ} = \Delta S + \Delta S_0 \geq 0. \qquad (5a)$$

This may be converted to an equality by introducing the so called entropy deficit function, $\theta$.[18-20]

$$\Delta S + \Delta S_0 - \theta = 0 \tag{5b}$$

Although the re-expression of eq. (5a) as eq. (5b) begins as no more than a book-keeping convenience – allowing the manipulation of equalities rather inequalities – it proves to be a remarkably advantageous starting point for a variety of derivation.[18,20] Notice that the above expression implies that $\theta = \Delta S_{Univ}$ and so $\theta$ *is equivalent to the excess entropy produced in the universe as the result of a given irreversible process* (appropriately averaged over all microstate realizations of the process). The primary aim of the present work is to derive a general expression for $\theta$ in terms of experimentally measurable quantities pertaining to an irreversible process.

We begin by considering a particular infinitesimal processes which produces a given amount of work, $\delta w$, and heat, $\delta q$, with the goal of quantifying the average excess entropy resulting from the associated mechanically and/or thermally irreversible mechanisms. We assume that prior to the infinitesimal process the system is in a state of equilibrium with a bath at temperature T. With little loss in generality, we further assume that each infinitesimal constraint displacement, *dX*, is carried out sufficiently rapidly that the process remains effectively adiabatic during the time over which the resulting work is exchanged. Notice that after any such constraint displacement has ceased, it is necessarily the case that no further work can be exchanged. At this point in the infinitesimal process the system may (or may not) have a well-defined temperature, and even if it does, that temperature may (or may not) be the same as that of the bath. Thus, coupling the system to the bath (without allowing any further constraint displacement) may in general produce some exchange of heat, $\delta q$, between the system and bath.

Since we assume that the mechanical portion of the infinitesimal process is performed adiabatically, Jarzynski's irreversible work theorem[3,5] may be used to related $\delta W = \langle \delta w \rangle$ to $\overline{\delta W} = \langle \overline{\delta w} \rangle$, and thus determine $\delta W_{diss}$. More specifically, Jarzynski's theorem demonstrates that the work performed in any adiabatic process (whether infinitesimal or not) may be related to the Helmholtz free energy change of the system (associated with the corresponding constraint displacement). When applied to an infinitesimal adiabatic process the Jarzynski theorem becomes

$$\langle e^{-\beta \delta w} \rangle = e^{-\beta dA} = e^{-\beta \langle \overline{\delta w} \rangle} = e^{-\beta \overline{\delta W}}. \tag{6a}$$

Notice that the brackets $\langle ... \rangle$ in the above equation are defined exactly as in eqs 2 and 3. The above result may clearly also be expressed as

$$\overline{\delta W} = \langle \overline{\delta w} \rangle = -\frac{1}{\beta} \ln \langle e^{-\beta \delta w} \rangle = -kT \ln \langle e^{-\beta \delta w} \rangle . \tag{6b}$$

Thus, the associated dissipative work is

$$\delta W_{diss} = \delta W - \overline{\delta W} = \langle \delta w \rangle + kT \ln \langle e^{-\beta \delta w} \rangle. \tag{7}$$

The first and second laws, eqs 4 and 7, may now be combined to obtain

$$\delta \theta = dS + dS_0 = \frac{\delta Q}{T} - \frac{\delta \overline{Q}}{T} = \frac{\delta \overline{Q} + \delta W - \overline{\delta W}}{T} - \frac{\delta \overline{Q}}{T}$$

$$= \frac{\delta W_{diss}}{T} = \frac{\langle \delta w \rangle}{T} + k \ln \langle e^{-\beta \delta w} \rangle \tag{8}$$

The above expressions implicitly assume that the temperature of the bath is maintained at a constant value of $T_0 = T$ throughout the processes. Under these conditions, the mechanical irreversibility, and the resulting dissipative work, leads to entropy production. The above expressions also imply that any (transient) change in the temperature of the system which is produced as a consequence of such an inifinitesimal process, is also infinitesimal and so does not have a first order effect on the excess entropy. In other words, if we assume that the effective temperature of the system at the end of the adiabatic work exchange is changed by $\delta T$ then the above entropy increase may be expressed as $\delta\theta = dS + dS_0 = \frac{\delta Q}{T} - \frac{\delta Q}{T + \delta T} \approx \frac{\delta Q}{T} - \frac{\delta Q}{T} + O(\delta Q \times \delta T)$ which is clearly equivalent to eq 8, to first differential order.

More generally, we may envision a process in which the system is initially equilibrated with a bath at temperature $T$, but is coupled to a different bath of temperature $T_0$ after the irreversible work exchange. In this case eq. 8 becomes

$$\delta\theta = dS + dS_0 = \frac{\delta Q}{T} - \frac{\delta Q}{T_0} = \frac{\delta W_{diss}}{T} + \left(\frac{1}{T} - \frac{1}{T_0}\right)\delta Q. \qquad (9a)$$

Notice that the heat exchange, $\delta Q = dU - \delta W$, is again determined by the first law, where $dU$ is now the equilibrium energy change associated with both changing the constraint(s) on the system *and* changing the system temperature from $T$ to $T_0$. By combining the above result with eq. 7 we obtain,

$$\delta\theta = \frac{\langle\delta w\rangle}{T} + k\ln\langle e^{-\beta\delta w}\rangle + \left(\frac{1}{T} - \frac{1}{T_0}\right)\langle\delta q\rangle. \qquad (9b)$$

The last term on the right-hand-side reflects the entropy increase produced as a result of the *thermally* irreversible exchange of heat between the system (at temperature $T$) and the bath (at temperature $T_0$). Note that if $T_0$ only differ from $T$ by an infinitesimal amount then the resulting entropy increase again becomes equivalent to eq 8, to first differential order. In other words, *no thermally irreversible entropy production occurs to first order in the temperature difference between the system and bath*. This may be easily confirmed by specific examples, such as the cooling (or heating) of copper block of temperature $T$ by an ideal gas temperature bath of temperature $T_0$, in which case excess entropy is only produced to second order in $\Delta T = T_0 - T$.[20]

Notice that the above results place no restriction on the magnitude of $\Delta T$, or the nature of the constraint(s) along which work is performed. Moreover, although most of the above expressions pertain to infinitesimal processes they may be applied to a wide variety of non-infinitesimal process (as further discussed in Section 5).

## 4. Quasi-Static Irreversible Processes

The general results described in Section 3 require measuring appropriately averaged irreversible work and heat exchanges in order to quantify excess entropy production. Since the time-dependent generalized force, $f(t)$, occurring in the course of such work exchanges need not track any thermodynamic state function of either the system or surrounding, it is not in general possible to express $W$, and thus also $Q$ and $\theta$, in terms of such state functions. However, one may identify a broad class of quasi-static irreversible (QSI) processes for which it is possible to quantify $W$, $Q$, and $\theta$, by directly integrating thermodynamic state functions.

We here define a QSI process as ones in which the system and bath each separately remain in states that are arbitrarily close to *internal equilibrium*, although they may deviate arbitrarily far from equilibrium with each other. In other words QSI processes are ones in which the generalized force(s), $f$, and constraint(s), $X$, track thermodynamic functions of either the system or surroundings. More specifically, if the constraints are taken to be the natural variables of the system such that $\Psi(\{X_i\})$ is the associated thermodynamic energy function, then for any quasistatic process the constraint variables are necessarily equal to the thermodynamic derivative, $f_i = \left(\frac{\partial \Psi}{\partial X_i}\right)_{X_j \neq X_i}$. For example, if $\Psi = U$ and $\{X_i\}=S, V, n_1, n_2, n_3\ldots$ then $\{f_i\} = T, -P, \mu_1, \mu_2, \mu_3\ldots$ (and thus the general derivative relation between $X_i$ and $f_i$ also pertains to $T$ and $S$). Similar relations hold between the generalized forces and constraints within the surroundings. However, it is important to stress, again, that *the generalized forces of the system need not be in equilibrium with those in the surroundings*. Thus, for a QSI process involving irreversible pressure-volume work we may obtain the corresponding QSI work by integrating $W_{QSI} = -\int P_0 dV$. Notice that this requires integrating the pressure of the *bath* over the volume change of the *system*. More generally, we will establish the conditions under which all other kinds of irreversible QSI work exchanges may be quantified using $W_{QSI} = \sum_i \int (f_i)_0 dX_i$.

One may imagine various practical realizations of QSI processes. For example, we may envision a system at some high temperature which is weakly coupled to a bath at some lower temperature, such that the system retains a well defined temperature (and equation of state) as it slowly cools by exchanging heat with the bath (which may in turn be assumed to be sufficiently large that its temperature remains effectively constant). Alternatively, one may envision a high pressure gas tank surrounded by a bath at a lower pressure. If the volume of the gas is constrained by a piston which is only allowed to move in small steps (for example, by removing a series of stops) then the system may again remain arbitrarily close to internal equilibrium, such that the irreversible work done on the surroundings is the product of the system volume change times the pressure of the surroundings. One may also imagine approximating much more complicated phenomena as QSI processes, such as, for example, a child snorkeling in the sea while maintaining more-or-less well defined subcutaneous values of $T$ and $\mu_i$ that differ from those of the surrounding aqueous solution.

A key difference between QSI and more general irreversible processes is that we may apply the fundamental equation of thermodynamics independently to both the system and surroundings, and so for any infinitesimal QSI process

$$dU = TdS + \sum_i f_i dX_i \quad (10a)$$

$$dU_0 = T_0 dS_0 + \sum_i (f_i)_0 (dX_i)_0 . \quad (10b)$$

Energy conservation further requires that $dU = -dU_0$, and so the energy change of the system may also be expressed as,

$$dU = -T_0 dS_0 - \sum_i (f_i)_0 (dX_i)_0 , \quad (10c)$$

or eqs 10a and 10c may be equated to obtain,

$$TdS + T_0 dS_0 = -\left[\sum_i f_i dX_i + \sum_i (f_i)_0 (dX_i)_0 \right]. \quad (11)$$

Moreover, the second law requires that $dS_0 = \delta\theta - dS$ which, when combined with the eq 11, implies the following general expression for the entropy produced in any QSI process.

$$\delta\theta_{QSI} = \left(\frac{T_0 - T}{T_0}\right)dS - \frac{1}{T_0}\left[\sum_i f_i dX_i + \sum_i (f_i)_0 (dX_i)_0\right] \qquad (12)$$

With little loss of generality we may further assume that the surroundings (bath) is sufficiently large and well mixed that all of it behaves ideally, and so $T_0$, and all of the $(f_i)_0$ – such as the $P_0$ and the chemical potentials (concentrations) of every type of particle in the bath – are all constant. This ideal bath approximation is implicitly assumed throughout the foregoing analysis (although there is no fundamental difficulty associated with relaxing this approximation, for example, by assuming that the bath is finite and so its intensive variables may vary as a result of heat and work exchanges with the system).

We may further specify global condition(s) which establish a quantitative connection between changes in the extensive variables of system and bath. For example, we may assume that all of the extensive thermodynamic constraints, $X_i$, are strictly conserved in the sense that the total values of each $X_i$ variable is constant (from the perspective of the entire universe), such that $dX_i = -(dX_i)_0$. For example, if $X_i$ represents a volume constraint then this conservation condition implies that the volume of the universe is constant, while if $X_i$ represents the number particles of a particular type, then the total number of these particles in the universe is fixed. Under such strictly conservative quasistatic irreversible (C-QSI) conditions, the above expressions may be combined to produce the following additional identity.

$$dU = T_0 dS + \sum_i (f_i)_0 dX_i - T_0 \delta\theta_{C-QSI}. \qquad (13)$$

Note that the first two quantities on the right-hand-side of the above equation represent thermodynamic expressions for the C-QSI heat and generalized work exchanges, respectively. Thus, $\sum_i \int (f_i)_0 dX_i$ represents a general expression for the irreversible work exchanged in any C-QSI process.

Equation 12, plus the global conservation condition, imply that the net entropy produced in any C-QSI process is

$$\delta\theta_{C-QSI} = \left(\frac{T_0 - T}{T_0}\right) dS + \sum_i \left(\frac{(f_i)_0 - f_i}{T_0}\right) dX_i . \qquad (14)$$

Notice that this expression relates the irreversibly produced entropy in the *universe* to changes in the *system* extensive variables, $S$ and $X_i$, along a given QSI irreversible path. However, one must also know the state of the bath, as expressed by the values of its intensive variables, $T_0$ and $(f_i)_0$.

We can exchange any one (or more) of the above independent variables using the appropriate Legendre transformation. For example, we may swap $S$ and $T$ as independent variables by transforming the thermodynamic potential from $U(S, X_i)$ to $A(T, X_i) = U - \left(\frac{\partial U}{\partial T}\right)_{X_i} T = U - TS$. Thus, by substituting $dU = dA + TdS + SdT$ and $dU_0 = dA_0 + T_0 dS_0 + S_0 dT_0 = dA_0 + T_0 dS + S_0 dT - T_0 \delta\theta$ in eqs 10, we obtain the following expression for the Helmholtz free energy change of the system in any such C-QSI process.

$$dA = (T_0 - T) dS - SdT + \sum_i (f_i)_0 dX_i - T_0 \delta\theta_{C-QSI} \qquad (15a)$$

However, since $A(T, X_i)$ we should also explicitly express $S$ as a function $T$ and $X_i$ to obtain $dS = \left(\frac{\partial S}{\partial T}\right)_{X_i} dT + \sum_i \left[\left(\frac{\partial S}{\partial X_j}\right)_{T, X_j \neq X_i} dX_j\right]$ and thus,

$$dA = (T_0 - T)\left\{\left(\frac{\partial S}{\partial T}\right)_{X_i} dT + \sum_i \left[\left(\frac{\partial S}{\partial X_i}\right)_{T, X_j \neq X_i} dX_i\right]\right\} \quad (15b)$$
$$- SdT + \sum_i (f_i)_0 dX_i - T_0 \delta\theta_{C-QSI}$$

Moreover, since $A$ is a state function, we know that $dA = -SdT - \sum_i f_i dX_i$, which represents the free energy change obtained by following a reversible path between the same initial and final states of the system. By subtracting the above two expressions for $dA$, we obtain the following alternative expression for the C-QSI excess entropy, in terms of the new independent variable, $T$.

$$\delta\theta_{C-QSI} = \left(\frac{T_0 - T}{T_0}\right)\left\{\left(\frac{\partial S}{\partial T}\right)_{X_i} dT + \sum_i \left[\left(\frac{\partial S}{\partial X_i}\right)_{T, X_j \neq X_i} dX_i\right]\right\} + \sum_i \left(\frac{f_i - (f_i)_0}{T_0}\right) dX_i. \quad (16a)$$

In order to further quantify the above result, it is useful to consider processes for which the generalized displacements coordinates represent the volume and numbers of molecules of each component of the system, $X_1 = V$ and $X_{j>1} = n_j$, and so the corresponding forces are, $f_1 = -P$, $(f_1)_0 = -P_0$, $f_{j>1} = \mu_j$, and $(f_{j>1})_0 = \mu_{j0}$ (where $P$ is pressure and $\mu_j$ is the chemical potential of component $j$). With these identifications, we may employ standard thermodynamic relations to re-express the above equation as follows.

$$\delta\theta_{C-QSI} = \left(\frac{T_0 - T}{T_0}\right)\left\{\left(\frac{C_V}{T}\right) dT + \left(\frac{\alpha_P}{\kappa_T}\right) dV + \sum_i s_i |_V \, dn_i\right\}$$
$$- \left(\frac{P_0 - P}{T_0}\right) dV + \sum_i \left(\frac{\mu_{i0} - \mu_i}{T_0}\right) dn_i \quad (16b)$$

The standard thermodynamics parameters appearing in the above expression are

$$C_V = T\left(\frac{\partial S}{\partial T}\right)_{V, n_i}, \quad \alpha_P = \frac{1}{V}\left(\frac{\partial V}{\partial T}\right)_{V, n_i}, \quad \kappa_T = -\frac{1}{V}\left(\frac{\partial V}{\partial P}\right)_{T, n_i}, \quad \text{and}$$

$$s_j\,|_V = \left(\frac{\partial S}{\partial n_j}\right)_{T,V,n_i \neq n_j} = -\left(\frac{\partial \mu_j}{\partial T}\right)_{V,n_i}.$$ Notice that for a *reversible* process, the intensive variables of the system and surroundings are necessarily equal, $T = T_0$, $P=P_0$ and $\mu_j=\mu_{j0}$, and so the above expression implies that no entropy is produced, as expected.

Clearly the above procedure may be repeated by considering other thermodynamic potential functions, using the appropriate Legendre tranformations. A particularly important example is Gibbs free energy, $G(T,P,n_i) = U - TS + PV$ whose total differential is $dG = dU - TdS - SdT + PdV + VdP$. Proceeding as above, we obtain the following expressions for $dG$ and $\delta\theta$ for a C-QSI process in which $T$, $P$ and $n_i$ are treated as independent variables.

$$dG = (T_0 - T)\left[\left(\frac{C_P}{T}\right)dT - \alpha_P V dP + \sum_i \bar{s}_i dn_i\right]$$
$$-(P_0 - P)\left[\alpha_P V dT - \kappa_T V dP + \sum_i \bar{v}_i dn_i\right] - SdT + VdP + \sum_i \mu_{i0} dn_i - T_0 \delta\theta_{C-QSI} \tag{17}$$

$$\delta\theta_{C-QSI} = \left(\frac{T_0 - T}{T_0}\right)\left[\left(\frac{C_P}{T}\right)dT - \alpha_P V dP + \sum_i \bar{s}_i dn_i\right]$$
$$-\left(\frac{P_0 - P}{T_0}\right)\left[\alpha_P V dT - \kappa_T V dP + \sum_i \bar{v}_i dn_i\right] + \sum_i\left(\frac{\mu_{i0} - \mu_i}{T_0}\right)dn_i \tag{18}$$

The partial molar quantities appearing in the above expressions are defined, as usual, by

$$\bar{s}_j = \left(\frac{\partial S}{\partial n_j}\right)_{T,P,n_i \neq n_j} = -\left(\frac{\partial \mu_j}{\partial T}\right)_{P,n_i} \text{ and } \bar{v}_j = \left(\frac{\partial V}{\partial n_j}\right)_{T,P,n_i \neq n_j} = \left(\frac{\partial \mu_j}{\partial P}\right)_{T,n_i}.$$

The above examples should suffice to illustrate how any number of other Legendre transformations (both standard and non-standard) could be implemented to convert eqs 13 and 14 to functions of other independent variables. Moreover, the generalized constraints, $X_i$, in eqs 13 and 14 are clearly not limited to the variables $V$ and $n_i$, as one could readily include constraints pertaining to the work performed against an external field (e.g. magnetic or gravitational etc.) or interfacial phenomena (e.g. energy changes associated with the creation of surface or line contacts etc.).

## 4. Discussion and Conclusions

The Jarzynski theorem, upon which our most general results are founded, pertains to irreversible work performed on any Hamiltonian dynamical systems. In other words, the theorem as originally derived by Jarzynski pertains to adiabatic processes. However, the second law results in section 3 are certainly not restricted to adiabatic processes, as these allow heat exchanges to take place (after all constraint displacements have ceased). Thus, the present results appear to be restricted to processes in which work and heat exchanges occur at different times. However, this restriction is more apparent than real, as several studies have demonstrated that the Jarzynski theorem is in fact more generally applicable. For example, Crooks has shown[15] that the Jarzynski theorem also holds for any system governed by stochastic Markovian (microscopically reversible) dynamics, such as one whose properties are determined using a strictly thermostated Metropolis Monte Carlo simulation algorithm.[6,21] Thus, the Jarzynski theorem is applicable to both adiabatic and isothermal processes, and so also presumably to any hybrids of these two extremes.

Although most of the expressions in Section 3 pertain to infinitesimal processes, these may readily be applied to a wide range of non-infinitesimal irreversible process. For example, for any non-infinitesimal process in which all the work is performed adiabatically, followed by re-coupling the system to a bath (of either temperature $T$ or $T_0$), eqs 8 and 9 may be applied directly to represent the associated non-infinitesimal dissipative work and heat exchanges, and the resulting mechanical and/or thermal entropy production. Other types of non-infinitesimal irreversible process could be approximated by an appropriately defined sequence of infinitesimal processes. For example, a partially adiabatic processes could be approximated by allowing heat exchange to occur only in some fraction of the infinitesimal irreversible steps used to represent the process. In general, one could introduced baths of various intermediate temperatures along the course of such a sequence of infinitesimal processes in order to allow sufficient flexibility to accurately represent the time dependent work and heat exchanges in a particular process of interest. However, it is important to note that in applying eq 9b one must assume that the system has relaxed to a state of internal equilibrium at the end of each step. Thus, eq 9b may be summed over an arbitrary number of such infinitesimal (or small finite) steps to produce the following rectified form of the second law of thermodynamic, which is applicable to a wide variety of non-infinitesimal irreversible processes.

$$\theta = \sum_l \left(\frac{\langle \delta w \rangle}{T}\right)_l + k \sum_l \ln \langle e^{-\beta \delta w} \rangle_l + \sum_l \left(\frac{1}{T} - \frac{1}{T_0}\right)_l \langle \delta q \rangle_l \qquad (19)$$

Notice that all of the variables on the right-hand-side of eq 19 are experimentally quantifiable. Thus, eq 19 rectifies the second law to facilitate the experimental measurement of the excess entropy changes, just as the first law facilitates the measurement of energy changes. Stated in another way, eq 19 quantifies the entropy produced as the result of *irreversible* work and heat exchanges, just as the eq 1 quantifies the energy produced by *reversible* work and heat exchanges. Moreover, both expressions may be applied to converse situations, in the sense that eq 19 is equally applicable to reversible processes (in which case it predicts that $\theta=0$) while eq 1 also holds for irreversible processes (since $\Delta U$ is path independent).

Although the Jarzynski theorem is quite general, statistical considerations limit its practical application to processes in which the dissipative work is not much larger than kT (or RT in molar units).[3,5] For processes involving greater dissipative work exchange, the Jarzynski theorem still holds, but measuring the required average irreversible work becomes difficult, as this average becomes increasingly dominated by rare realizations.[5] For molecular processes, such as RNA pulling experiments, dissipative work can be accurately measured.[7,8,17] Thus, one may readily quantify the mechanical and/or thermal excess entropy produced in such processes, as further discussed below.

The QSI results presented in Section 4 are not restricted with regard to system size or the magnitude of the disparities between the properties of the system and bath. However, both the system and bath are assumed to remain continuously in states with well defined intensive properties (i.e. $T$, $P$ and $\mu_i$). Thus any QSI process may be fully described simply by specifying the path over which constraints are displaced, as these displacements now track equilibrium thermodynamic variables, and so uniquely specify the state of the system (and bath) at every point along the path. Thus, the expressions in Section 4 can readily be integrated over any such path. For example, eq 18 may be integrated to obtain the following expression may be used to evaluate the entropy produced in any C-QSI process, with $T$ and $P$, and $n_i$ treated as independent system variables.

$$\theta_{C-QSI} = \left[ \int \left( \frac{C_P}{T} \right) dT - \int \alpha_P V dP + \sum_i \int \bar{s}_i dn_i \right] - \frac{1}{T_0} \left[ \int C_P dT - \int T\alpha_P V dP + \sum_i \int T\bar{s}_i dn_i \right]$$

$$- \frac{P_0}{T_0} \left[ \int \alpha_P V dT - \int \kappa_T V dP + \sum_i \int \bar{v}_i dn_i \right] + \frac{1}{T_0} \left[ \int \alpha_P P V dT - \int \kappa_T P V dP + \sum_i \int P \bar{v}_i dn_i \right] \quad (20)$$

$$+ \frac{1}{T_0} \left[ \sum_i \mu_{i0} n_i - \sum_i \int \mu_i dn_i \right]$$

Each of the integrals in the above expression are carried out over the equilibrium trajectory of the system defined by the associated QSI path, which may in general involve arbitrarily large excursions in the $T$, $P$ and $n_i$ values of the system (while the bath is assumed to be ideal, so its intensive variables are effectively constant).

Notice that the QSI approximation is less restrictive than the approximation usually invoked in theories of irreversible thermodynamics. More specifically, the QSI approximation places no restrictions on the magnitude of the imbalance between the system and bath intensive variables, $T_0 - T$, $P_0 - P$ and $\mu_{0i} - \mu_i$ or more generally, $(f_i)_0 - f_i$. If we had assumed that these differences were small, then we could have performed Taylor expansions in terms these differentials, leading to first order terms in the usual irreversible thermodynamic expansion, and higher order terms representing the associated fluctuations. However, no such expansion is required in order to quantify the entropy produced in any QSI process.

As a representative application of the above results, consider the excess entropy produced in the irreversible unfolding of an RNA (P5abc domain).[8] Bustamante and co-workers have experimentally verified the applicability of the Jarzynski theorem to this process, by obtaining an unfolding free energy of ΔG ~ 60 kT from both reversible and irreversible work measurements. In a particular irreversible process a dissipative work of ~3 kT was measured, and thus eq 9a implies that an excess entropy of ~25 J/(K mol) is produced, assuming that the RNA molecule remains at ambient temperature throughout the unfolding process. However, since the enthalpy of unfolding of the RNA is ΔH ~ 130 kT (in 200 mM NaCl and 10 mM Na phosphate buffer at pH 6.4)[22] one would expect significant heat exchange to accompany the unfolding process. If the unfolding were performed sufficiently slowly that the RNA remains isothermal, then the heat exchange would be fully reversible, and so would not produce any additional excess entropy. However, if the extension were carried out adiabatically then the temperature of the RNA would decrease upon unfolding, by an amount that depends on its heat capacity. If we further assume that such an adiabatic unfolding takes place for an RNA molecule that is hydrated by 10 water molecules per RNA base (and thus a total of 690 waters hydrating the entire 69 base RNA domain), and that the heat capacity of the hydrated RNA may be approximated by that of bulk water (~75 J/K per mole of water) then the above ΔH implies that the RNA would cool by about 6 K as the result of unfolding. If we further assume that this hydrated RNA is weakly coupled weakly to a bath at ambient temperature (i.e. a rare gas at 298K) then we expect the temperature of RNA to remain well defined as it slowly equilibrates from 292 K to 298K subsequent to the pulling experiment (with no further constraint displacement). Under these conditions we may evaluate the temperature integrals in eq 20 to obtain $C_P[\ln(1+\Delta) - X/(1+\Delta)] \approx C_P \Delta^2 / 2$ for the thermally irreversible excess entropy,[20] where $\Delta = \Delta T/T_0 = -6/298 = -0.02$, and $C_P$ ~ 330 kJ/(K mol) is the total heat capacity of the hydrated RNA complex. Under such conditions, we thus estimate that ~10 J(K mol) of excess entropy would be produced solely as the result of the thermally irreversible heat transfer. Therefore, the total mechanical and thermal excess entropy produced in the universe as the result such an irreversible process would be ~35 J/(K mol) = 25 J/(K mol) + 10 J/(K mol).

In summary, we have demonstrated that the second law of thermodynamics may be rectified to obtain equalities rather than inequalities pertaining to irreversible entropy production. We have derived general equations for entropy production in terms of irreversible work and heat transfers from the Jarzynski theorom. Although these expressions are quite general, practical limitations restrict their application to molecular or mesoscopic processes.[5] These restrictions do not apply to QSI processes, for which one may calculate irreversible entropy production in terms of standard thermodynamic variables of the system (as well as the values of the intensive variables of the bath).

ACKNOWLEDGMENTS

This work was supported by the National Science Foundation (CHE-0455968). The authors thank Professor Igal Szleifer of Purdue University for insightful discussions.